\documentclass{article}
\usepackage{hiph-preprint}
\volnumber{22} \issuenumber{1} \edyear{2005}                             
\frompage{000} \topage{000}                                              
\recrevdate{1 January 2005}                                              
\def\rightmark{Proposal for a High Energy Nuclear Database}
\def\leftmark{D.A. Brown and R. Vogt}
 
\title{A High Energy Nuclear Database Proposal} 
\authors{ 
    {David A. Brown$^1$ and Ramona Vogt$^{2}$}\\[2.812mm]
    {
        \normalsize
        \hspace*{-8pt}$^1$ Lawrence Livermore National Laboratory,
        Livermore, CA, USA\\[0.2ex]
        \hspace*{-8pt}$^2$ Lawrence Berkeley National Laboratory,
        Berkeley, CA, USA \\ and \\
        Department of Physics, UC Davis, Davis, CA, USA
    }
}

\abstract{We propose to develop a high-energy heavy-ion experimental database 
and make it accessible to the scientific community through an on-line 
interface. This database will be searchable and cross-indexed with relevant 
publications, including published detector descriptions.  Since this database 
will be a community resource, it requires the high-energy nuclear physics 
community's financial and manpower support.  This database should eventually 
contain all published data from the 
Bevalac, AGS and SPS to RHIC and LHC energies 
proton-proton to nucleus-nucleus collisions as well as other relevant systems, 
and all measured observables.  Such a database would have tremendous 
scientific payoff as it makes systematic studies easier and allows simpler 
benchmarking of theoretical models to a broad range of old and new experiments.
Furthermore, there is a growing need for compilations of high-energy nuclear 
data for applications including stockpile stewardship, technology development 
for inertial confinement fusion and target and source development for upcoming 
facilities such as the Next Linear Collider.  To enhance the utility of this 
database, we propose periodically performing evaluations of the data and 
summarizing the results in topical reviews.
}

 
\makeindex
\begin{document}
 
\maketitle

\section{Background and Potential Impact}
 
We propose to create and maintain a high-energy nuclear database.  This 
central database will be web-accessible and searchable.   As with Evaluated 
Nuclear Data File (ENDF/B) and EXchange FORmat (EXFOR) databases \cite{bib1} 
and HEPDATA website \cite{bib2}, we will store cross sections, particle yields,
and single particle spectra.  We will also store data specific to higher 
energy reactions such as multi-particle spectra, flow and correlation 
observables.  Thus we seek to archive whatever is needed to 
characterize a high-energy heavy-ion reaction.  Initially we will focus on 
published measurements but eventually we will cross-link the data with 
experiment descriptions.  We also envision evaluating high-energy nuclear data 
and reporting these results in  periodic topical reviews of subsets of the 
data.  The idea of publishing the topical reviews has already sparked the 
interest of a few review journals.
 
The utility of such a database is clear: it would organize existing data, 
allowing easier cross-experiment comparisons, theory benchmarking and 
development of systematics.  In addition to the basic science needs, there 
is a growing list of applications for high energy nuclear data including:
understanding backgrounds in proton radiography; heavy-ion driven inertial
confinement fusion; $\nu$ and $\mu$ secondary beam source development for
MINOS and the Next Linear Collider; and cosmic ray dose rates for space
exploration.  Most applications do not use the data directly.  Instead, 
evaluated representations of the data are accessed by application codes.
 
Surprisingly, there is no national or international effort to collect and 
maintain such a database.  The US Nuclear Data Program (USNDP) \cite{bib1} 
has compiled low-energy nuclear reaction data for decades in the ENDF/B and 
other databases.  Similarly, the high-energy particle physics community is 
served by the HEPDATA, Particle Data Group (PDG), 
arXiv.org and SLAC-SPIRES websites.  
The high-energy nuclear physics community is only partially served by these 
data sources.  One could argue that most experiments make their published 
data available: for example, PHENIX posts tables of published data on the 
collaboration website.  Inevitably this leads to a proliferation of data 
formats and web sites.  Furthermore, experiments end and their web servers 
may no longer be maintained.  Thus, there is a very real risk that the data 
could be lost.  Given the volumes of data generated by experiments at the RHIC 
and future experiments at the LHC, GSI and elsewhere, this oversight should 
be rectified.
                                                                  
In the next few sections, we explain that the proposed database should be a 
community effort, motivate the need for data evaluation and topical reviews 
and provide some technical details.  We conclude with a status of the 
database proposal.  
 
\section{Database Management Philosophy}
 
Since this database would be a community resource, we propose a community 
driven management model such as the arXiv.org preprint server: the 
``consumers'' of arXiv.org are also its ``suppliers.''   
Physicists submit their preprints to arXiv.org because it serves as a form 
of advertising.  They browse arXiv.org because they know 
others are submitting their latest results there.  In this way, data 
collection is farmed out to the data producers -- a tactic we wish to employ.
 
The proposed database would differ from arXiv.org in two key respects.  First, 
the proposed database would not only contain published data but also auxiliary 
or supporting data sets that may be too large for publications such as 
Physical Review Letters.  Second, in order to assure that we only have 
high-quality data, we would like to piggyback on various journals' peer-review 
processes.  Ideally this means that authors will submit both the published and 
auxiliary supporting data to the database when submitting papers for 
publication.  One submission model would be to add links directly to journal 
submission pages.  Preliminary discussions with journal editors indicate the 
willingness of the journals to cooperate in this endeavor.
 
For the experimental collaborations to have the political will to support this 
project, they must be given both a financial stake in the eventual product and 
have a hand in steering the database development.  To encourage this, we 
propose holding annual workshops to guide development as well as discuss the 
topical reviews and propose new subjects for review.
 
\section{Evaluations and Topical Reviews}\label{review}
 
Evaluated data provides our ``best guess'' representation of a particular 
observable.  Whether one obtains this evaluation through a 
model calculation, systematics or a fit to raw experimental data, the final 
product needs to be checked against existing data and peer 
reviewed.  However the evaluation is produced, we need a better understanding 
of the raw experimental data.  For example, while charm production is a hot
new topic in heavy ion physics, charm hadrons, particularly $D$ mesons, have
been measured in a number of experiments over a wide energy range.  The
extrapolation of these measurements to the total charm cross section vary 
considerably, even at the same energy, presumably since many of these
measurements were taken before next-to-leading order perturbative QCD 
calculations became available.  The measurements were also hampered by small
statistical samples.  To determine the consistency of these data with
new measurements at heavy-ion colliders, these data must be re-evaluated
with all previous assumptions re-examined.  A topical review would provide
a clear and systematic examination of all data and clarify the situation
considerably.

\section{Technical Details}
 
Many of the tools needed for this database are available 
``off-the-shelf.''  We envision that users will access the database through 
a set of JSP or PHP dynamic web pages.  Both JSP and PHP have tools that 
simplify on-line database queries.  The data itself will be stored in
a MySQL relation 
database.  We also envision that the dynamic web pages will be able to send 
the data directly to a plotting utility such as LLNL's LLNLPlot Java applet.  
LLNLPlot can plot 2D and 3D data and is the plotting back-end of the Nuclear 
and Atomic Data System (NADS) \cite{bib11}.
 
A central, yet often neglected, aspect of data archives is the technical 
details of the data storage format.  The nuclear data community traditionally 
suffers from a multitude of relatively obscure data formats. For example, the 
format still used in the ENDF/B database was designed specifically to 
accommodate the limitations of now-obsolete punch cards.  In some cases, the 
task of writing translation and visualization tools for these data sets 
requires a large, dedicated effort.
 
Given the importance of using a transparent and well-supported format, we have 
decided on XML (eXtensible Markup Language) as our data storage format. 
Documents stored in XML can be self-describing so that, with minimal effort, 
scientists/users 30 years from now can interpret the documents' contents.  
Furthermore, XML documents are represented by computationally convenient tree 
structures rather than the simple strings typically used to store nuclear 
data.  XML is a mature technology with the support of thousands of programmers 
and web developers and is extensively supported by most common programming 
languages.  Lastly, the many tools needed for web-based access and 
manipulation of XML databases have reached a state of maturity.
 
\section{Current Status}
 
We have submitted a white paper describing our proposal to the DOE-OS 
Heavy-Ion and Nuclear Theory programs and are circulating it in the STAR and 
PHENIX collaborations.  Copies of the whitepaper are available upon request 
from the authors.  Due to the budget cuts in FY05-06, it is unlikely to be 
funded before FY07.  Despite the funding uncertainties plaguing the field, 
elements of this proposed project are in the process of being developed for 
other uses, namely the XML data format and the LLNLPlot plotting tool.  Since 
we want this database to be a community resource, we strongly encourage 
members of the heavy-ion community to contact
us with their questions, comments, wishes and ideas.
 
\section*{Acknowledgments}
 
This work was performed under the auspices of the U.S. Department of Energy by 
University of California, Lawrence Livermore National Laboratory under 
Contract W-7405-Eng-48. R.V. was supported in part by the Director,
Office of Energy Research, Office of High Energy and Nuclear Physics,
Nuclear Physics Division of the U.S.\ Department of Energy
under Contract No.\ DE-AC02-05CH11231.

\vfill\eject
\end{document}